\documentclass[aps,prl,twocolumn,superscriptaddress,showpacs,floatfix]{revtex4}
\usepackage{amsmath,epsfig,subfigure}
\usepackage[active]{srcltx}
\usepackage{nicefrac}

\begin{document}
\title{Role of quantum nuclei and local fields in the x-ray absorption spectra of water and ice}
\author{Lingzhu Kong}
\affiliation{Department of Chemistry, Princeton University, Princeton, New Jersey 08544, USA}

\author{Xifan Wu}
\affiliation{Department of Physics,  Temple University, Philadelphia, Pennsylvania 19122, USA}

\author{Roberto Car}
\affiliation{Department of Chemistry and Department of Physics, Princeton University, Princeton, New Jersey 08544, USA}

\begin{abstract}
We calculate the x-ray absorption spectra of liquid water at ambient conditions and of hexagonal ice close
to melting, using a static GW approach that includes approximately local field effects. Quantum dynamics of the nuclei is taken 
into account by averaging the absorption cross section over molecular configurations generated by path 
integral simulations. We find that inclusion of quantum disorder is essential to bring the calculated 
spectra in close agreement with experiment. In particular, the intensity of the pre-edge feature, 
a spectral signature of broken and distorted hydrogen bonds, is accurately reproduced, in water and ice, 
only when quantum nuclei are considered. The effect of the local fields is less important but non 
negligible, particularly in ice. 
\end{abstract}

\pacs{61.25.Em, 61.05.cj, 71.15.Qe, 82.30.Rs}
\maketitle

In the last decades high resolution core level spectroscopy, such as x-ray absorption spectroscopy (XAS) 
or x-ray Raman scattering (XRS), has emerged as a powerful tool for the investigation of matter at the 
molecular scale~\cite{XAS}. Important studies of liquid water and other disordered hydrogen-bond (H-bond) 
environments have been performed with these techniques~\cite{Saykally2001,Wernet2004,Tse2008}. 
Core spectroscopy is  an element-specific local probe, complementary to diffraction for
structural information, but  requires a good theory of the excitation process  
as a prerequisite to a consistent experimental 
interpretation~\cite{Vinson2011,Vinson_PRB_2012}. 
Given that disordered environments need large cells and averages over different 
realizations, computationally efficient schemes based on density functional theory (DFT) have been the 
technique of choice in this context~\cite{xas-DFT}. DFT, however, is not an excitation theory and yields 
spectra that are at best in semi-quantitative agreement with experiment.

Dramatic improvement in the calculated spectra is obtained with a proper excitation theory, as shown in a 
recent paper~\cite{Wei2010}, which focused on water and adopted Hedin's GW approximation~\cite{Onida2002} 
for the self-energy of the excited electron in presence of a static core hole. Computationally, this 
approach is significantly more demanding than DFT and Ref.~\onlinecite{Wei2010} compromised on the 
structural model and the screening approximation. The calculated spectrum deviated from experiment mostly 
in the pre-edge intensity which was underestimated by $\sim$50\%.  
It was unclear if this discrepancy was due to limitations of the structural 
model, the screening model, or both. Ref.~\onlinecite{Wei2010} used liquid structures generated by 
{\sl ab initio} molecular dynamics (AIMD)~\cite{Car1985} in a periodic cell with 32 H$_2$O molecules. 
With standard DFT approximations and classical nuclear dynamics this scheme yields enhanced water 
structure. Thus the temperature of the simulation was set to $T\sim360K$ to improve agreement with 
experiment at ambient conditions ($T=300K$). In addition, a simple uniform screening model was adopted, 
neglecting the microscopic inhomogeneity of the medium. The same paper reported also the spectrum of a 
proton ordered cubic ice structure that minimized the DFT energy at $T=0K$. The deviation from experiment 
was significantly larger than in water, but one should recall that in real cubic (Ic) or hexagonal (Ih) ices, 
the distribution of the protons in the H-bonds is disordered. 
Moreover, the effect of zero-point motion was ignored. This is large, reflecting the quantum character of 
the nuclei, particularly the protons, signaled by neutron scattering experiments ~\cite{neutron}.

In this paper we address the above issues. We take the quantum character of the nuclei into account 
using molecular structures generated by path integral {\sl ab initio} molecular dynamics 
(PI-AIMD)~\cite{Marx1994} for water at $T=300K$ and for ice at $T=269K$~\cite{Morrone2008}. In addition, 
we improve the treatment of screening by including, albeit approximately, the local fields due to the
inhomogeneity of the medium within the Hybertsen-Louie (HL) ansatz~\cite{HL1988}. We adopt larger simulation 
cells than in Ref.~\onlinecite{Wei2010} and reduce the statistical error in the averages.  
Quantum effects bring the pre-edge intensity in close agreement with 
experiment in both water and ice. In addition, they reduce the post-edge intensity, for a better agreement 
with experiment, especially in water. Local field effects are modest in water, but cannot be neglected in 
ice that has a more open and less homogeneous microscopic structure. Here they improve the spectra in the 
main region that includes near- and post-edge, but some residual discrepancies remain, suggesting that a more accurate 
screening model could further improve the experimental comparison. Even at 
this stage, however, our approach, combining accurate {\sl ab initio} simulations and spectral calculations, 
is a powerful interpretive tool for x-ray spectroscopy.

We base our calculations on the molecular structures of Ref.~\onlinecite{Morrone2008}, 
which were obtained with PI-AIMD simulations using the BLYP functional~\cite{Becke1988,Lee2010} for the  
electronic energy within DFT. The corresponding classical AIMD structures with the same functional 
approximation and thermodynamic conditions were also reported in the same paper. 
Quantum nuclei soften the 
structure of water predicted by DFT, thereby greatly improving the experimental agreement of the 
calculated pair correlation functions. Residual overstructuring can be attributed to limitations of the 
functional approximation. Quantum dynamics increases the fraction of broken bonds and broadens the 
distribution functions. Thus, empirical adjustments like increasing the temperature of the simulation as in 
Ref.~\onlinecite{Wei2010} can be avoided. Ref.~\onlinecite{Morrone2008} used open PI 
trajectories~\cite{Morrone2007} to calculate the momentum distribution of the protons, finding excellent 
agreement with deep inelastic neutron scattering experiments. Relative to its classical counterpart, 
the quantum momentum distribution is strongly displaced towards higher momenta and deviates significantly 
from a spherical Gaussian distribution~\cite{Lin2011}, indicating that a higher effective temperature in 
classical simulations can only be a poor surrogate of full quantum simulations. The latter are based on 
Feynman paths in imaginary time, which provide an exact representation of equilibrium quantum statistical 
mechanics. The x-ray absorption cross section is a dynamic property requiring, in principle, 
path components in real time which would carry a complex phase not amenable to computer simulation, 
but we can rely on the wide separation of time scales between nuclear and electron dynamics, in the spirit 
of the Franck-Condon principle. In the present case it amounts to compute the spectra for the 
frozen nuclear configurations sampled with the adopted discrete representation of the Feynman paths in 
imaginary time.

Molecular configurations for water at $T=300 K$ and ice Ih at $T=269 K$ were generated in 
Ref.~\onlinecite{Morrone2008} by adopting the ``bulk" approximation 
in which the Feynman path of one proton per molecule is left open to improve the 
statistics of the momentum distribution~\cite{Morrone2007,Morrone2008}. It was shown that this approximation 
has negligible effect on the momentum distribution and very minor effect on the spatial correlations 
between the protons~\cite{Morrone2007}. The simulations used a periodic cell with 64 molecules for water 
and 96 molecules for ice Ih in a proton disordered configuration with cell volumes fixed to experiment. 
The Feynman paths were discretized using 32 replicas in both cases. In our XAS calculation we use the static 
Coulomb hole and screened exchange (COHSEX) approach of Ref.~\cite{Wei2010} and average the excitation cross 
section over all the molecules in the cell. We include quantum effects by averaging over path-integral replicas. 
Each particle ($H$ or $O$ nucleus) path runs in imaginary time from $\tau=0$ to
$\tau = \nicefrac{\hbar}{k_{B} T}$. To minimize the effect of the open paths we only include in the average 
26 middle path replicas in water and 16 in ice as this was sufficient for convergence. 
Even limiting the average to a fraction of the replicas, the number of calculations is huge 
and we consider only a single molecular dynamics snapshot to reduce the computational burden.. 
We checked for classical nuclei that this has minor effect on the spectra. Aiming at further reducing the 
computational cost, we have investigated whether calculations at the centroids of the paths could 
capture most quantum effects.  Unfortunately, this was not the case. 
In water, centroid calculations led to a pre-edge intensity approximately half-way between classical 
and quantum data, consistent with the observation that the fraction of broken bonds 
is approximately the same in the centroid and path structures.
In ice, that has no broken bonds, the centroid spectrum was indistinguishable from the classical one.

We include local fields in the screening by means of the HL local density approximation according 
to which the screened interaction reads~\cite{HL1988}
\begin{equation}
W(\mathbf{r},\mathbf{r'})=\frac{1}{2} \left ( W[\mathbf{r}-\mathbf{r'};\rho(\mathbf{r'})]+ \\
                                                          W[\mathbf{r'}-\mathbf{r};\rho(\mathbf{r})] \right )  
\end{equation}
Eq.(1) is based on the observation that the screening strength generally follows the local charge density. Here 
\begin{equation}
 W[\mathbf{r}' - \mathbf{r}; \rho (\mathbf{r}) ] = \frac{1}{(2\pi)^3}\int \epsilon^{-1}[\mathbf{q};\rho(\mathbf{r})] \\
v(\mathbf{q}) e^{i \mathbf{q} \cdot (\mathbf{r}-\mathbf{r^\prime})} \,d\mathbf{q}  \label{eq:screenedW}
\end{equation}
and $v$ is the bare Coulomb interaction. For the dielectric function, we adopt Bechstedt model~\cite{Bechstedt1992},
\begin{equation}
\epsilon[\mathbf{q},\rho(\mathbf{r})]=1+[(\epsilon _0-1)^{-1}+\alpha (\frac{q}{q_{TF}})^2+\frac{q^4}{4\omega_p^2}]^{-1}
\end{equation}
where q$_{TF}$ and $\omega_p$ denote Thomas-Fermi wavevector and plasmon frequency, respectively. 
These quantities depend on the local density but in Ref.~\cite{Wei2010} dependence on the average density was assumed. 
The $q=0$ value $\epsilon_0$ is taken from experiment and $\alpha$ is fixed by requiring  
Bechstedt model (Eq.(3)) to match the $q^2$ dependence of Penn model~\cite{Penn1962}. Eq.~\eqref{eq:screenedW}
can be analytically evaluated yielding
\begin{widetext}
\begin{equation}
W[\mathbf{r}' - \mathbf{r}; \rho (\mathbf{r}) ] = \frac{v(\mathbf{r}-\mathbf{r'})}{\epsilon_0} - \frac{1}{a(x_1-x_2)|\mathbf{r'
}-\mathbf{r}|}  \left ( \frac{e^{i (x_1)^{1/2}|\mathbf{r'}-\mathbf{r}|}}{x_1} - \frac{e^
{i (x_2)^{1/2}|\mathbf{r'}-\mathbf{r}|}}{x_2} \right ) 
\label{eq:integratedW}
\end{equation}
\end{widetext}
where  $x_{1,2} = (-b \pm \sqrt{b^2-4ac})/2a $ and $a=1/4\omega_p^2$, $b=\alpha/q_{TF}^2$ and $c=\epsilon_0/(\epsilon_0-1)$.
There are two contributions to the screened interaction $W$ in Eq.~(\ref{eq:integratedW}), the bare interaction $v$ 
divided by the macroscopic dielectric constant, and a local-density dependent screened interaction, which is 
fully nonlocal but short-ranged in space. To evaluate the effect of $W$ on the self-energy we use the 
convolution theorem for the part depending on the bare interaction, which is conveniently calculated in Fourier 
space, and we compute the integral in real space for the short-ranged nonlocal contribution.

\begin{figure}
{\epsfig{file=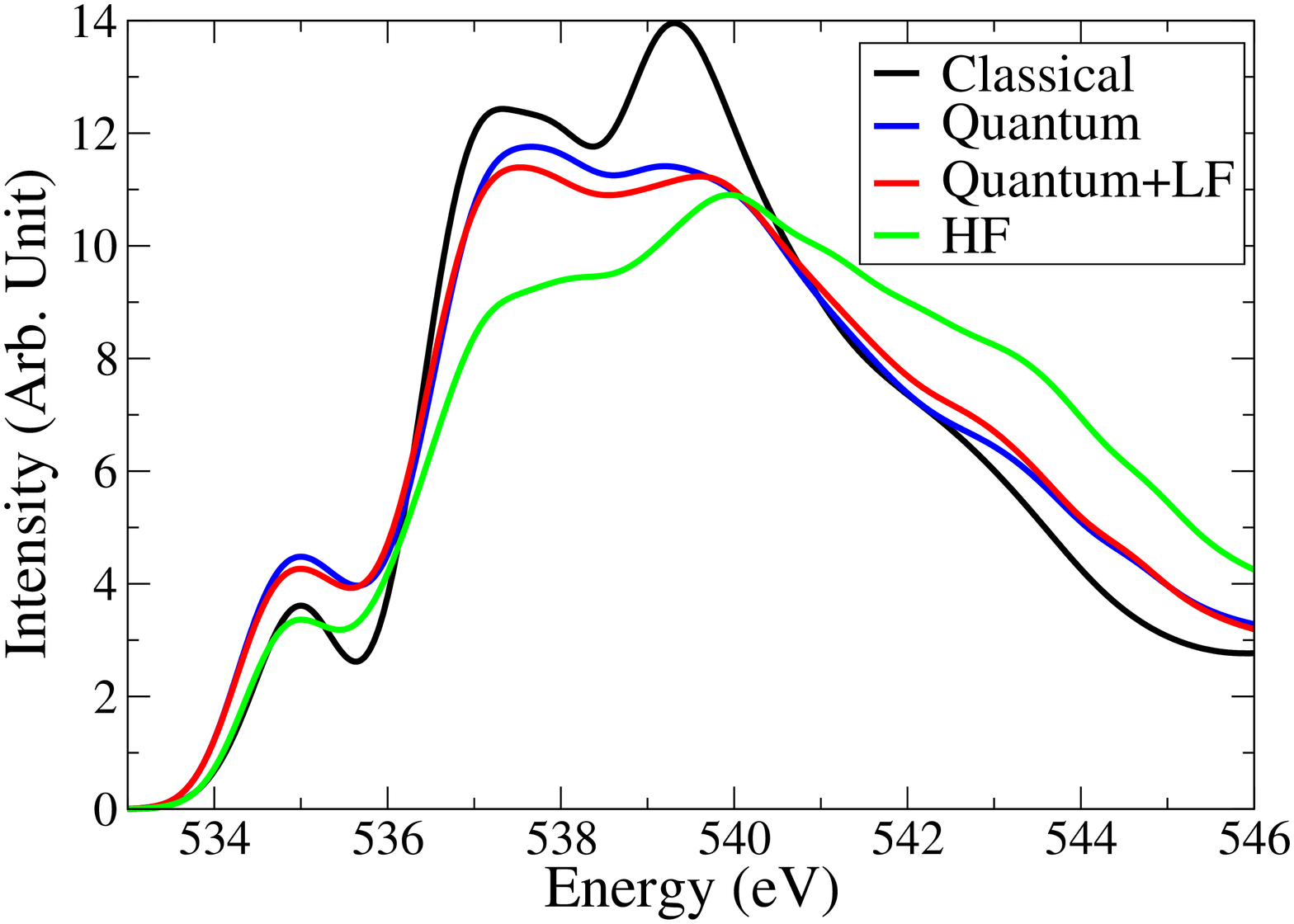,height=6cm,angle=270,clip=true}
\epsfig{file=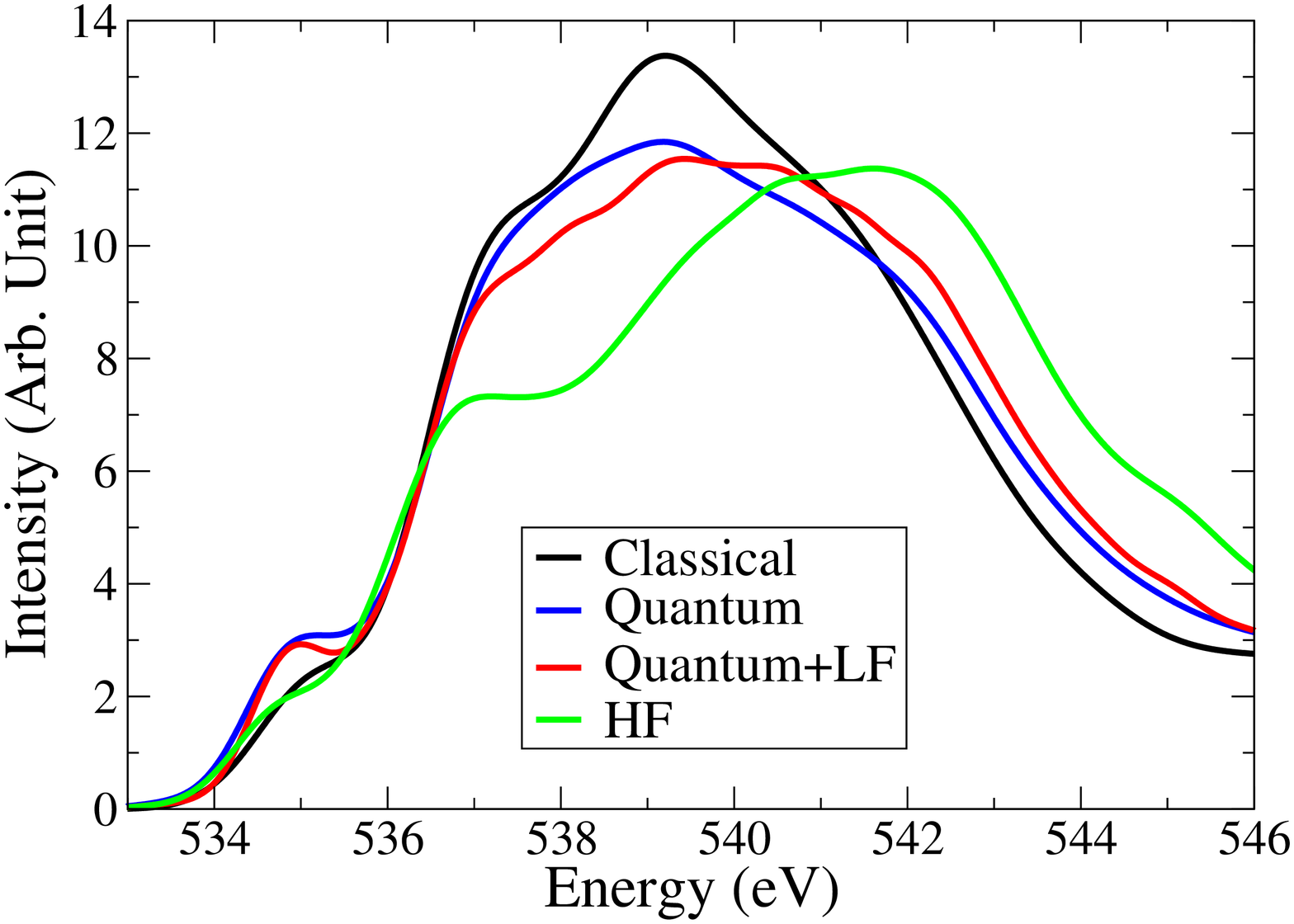,height=6cm,angle=270,clip=true} }
\caption{ (Color online) Calculated XAS spectra for liquid water at $T=300K$ (top) and ice Ih at $T=269K$ (bottom) 
according to four different schemes: (a) COHSEX with classical nuclei and uniform screening (black); (b) COHSEX 
with quantum nuclei and uniform screening (blue); (c) COHSEX with quantum nuclei and local field (LF) screening 
(see text) (red); (d) Hartree-Fock with quantum nuclei (green). The spectra are area normalized and Gaussian
broadened(fwhm=0.4 eV). In order of increasing energy, three features, the pre-edge, the near-edge, and the 
post-edge, characterize the spectra.}
\label{fig:xas_theory}
\end{figure}

Fig.~\ref{fig:xas_theory} shows the calculated XAS intensity for liquid water and ice Ih at different levels 
of approximation. The spectra are aligned at the onset and the absorption sum rule is imposed to fix the 
normalization so that the spectra have equal area in the energy range of the figure. Hartree-Fock results are 
also reported: they correspond to calculations with the dielectric function set equal to 1,
 i.e. ignoring screening.

Replacing classical with quantum nuclei affects the spectra, because zero-point motion 
significantly enhances the positional disorder relative to classical thermal motion. 
Qualitatively similar conclusions were reached in a recent study of the XAS spectrum of ice 
using PI configurations generated with an empirical intermolecular potential~\cite{Lee2010}.
In addition, quantum fluctuations increase the fraction of broken H-bonds in the liquid by $\sim$4\%~\cite{Morrone2008}. 
As a consequence pre-edge is increased, post-edge is reduced and there is an overall slight broadening 
of the spectrum. The effects on pre- and post-edge are more pronounced in the liquid, where the post-edge 
is particularly interesting as it correlates with the less overstructured H-bond network of the quantum simulations.

Local fields have comparatively smaller effects than quantum nuclei. Here it is useful to 
start from Hartree-Fock. In this limit screening is absent, the 
quasi-particle excitation energies are overestimated resulting in too wide spectra with the
post-edge more prominent than the near-edge in both water and ice. COHSEX calculations improve 
the agreement with experiment, by reducing the overall width and 
by lowering post-edge below near-edge in the liquid. The difference between the spectra with a uniform screening 
model and those obtained with the HL local density 
approximation is a measure of the inhomogeneity of matter at the molecular scale, which renders screening 
less effective than in the corresponding uniform medium. Local fields are underestimated in the HL 
approach~\cite{Onida2002} and one should expect that screening would be further reduced in a more accurate theory.
The effect should be small in water where the HL induced change is minor, but in ice the HL 
induced change is non negligible suggesting that a better screening model should move 
the spectrum further in the Hartree-Fock direction..

\begin{figure}
{\epsfig{file=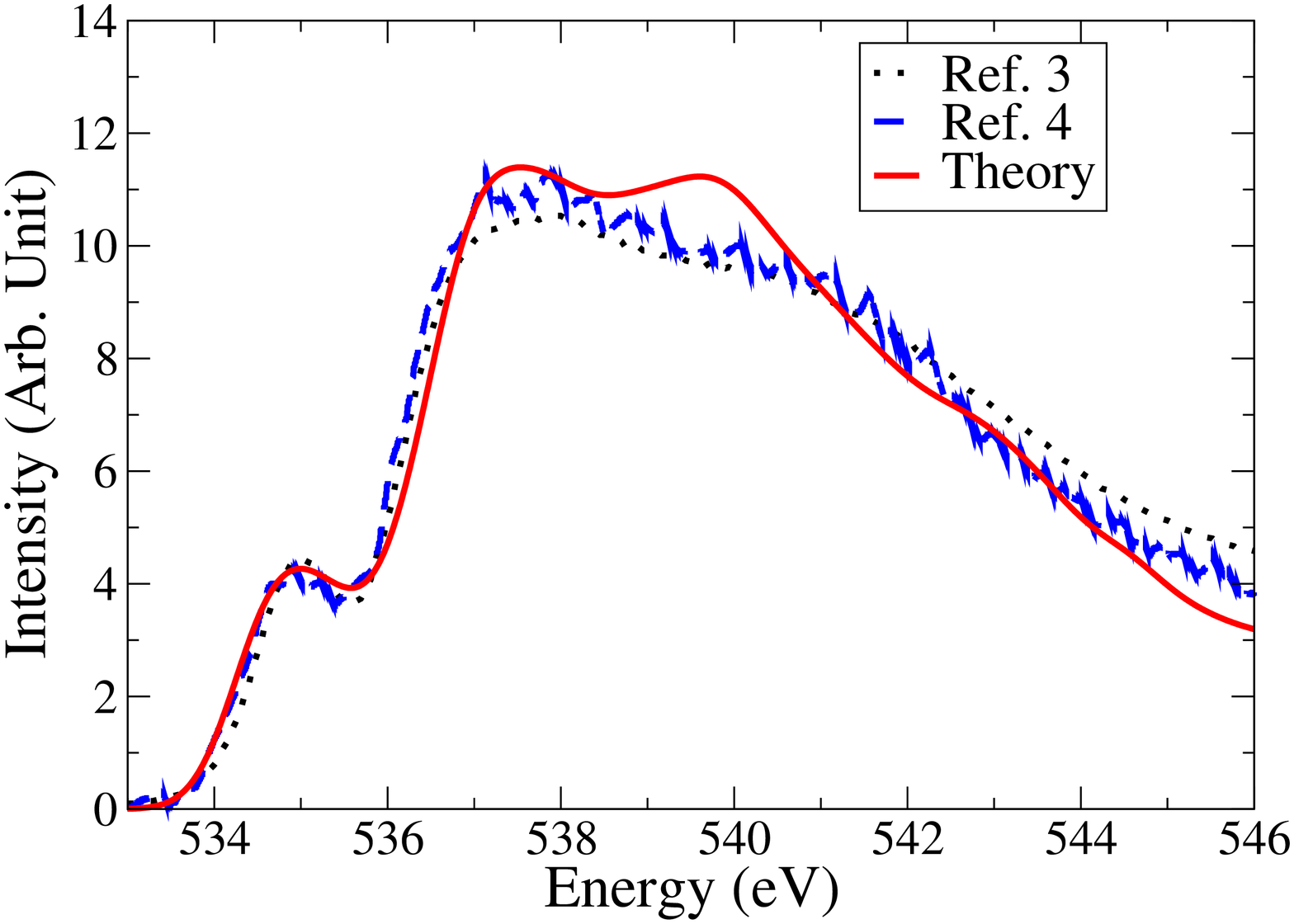,height=6cm,angle=270,clip=true}
 \epsfig{file=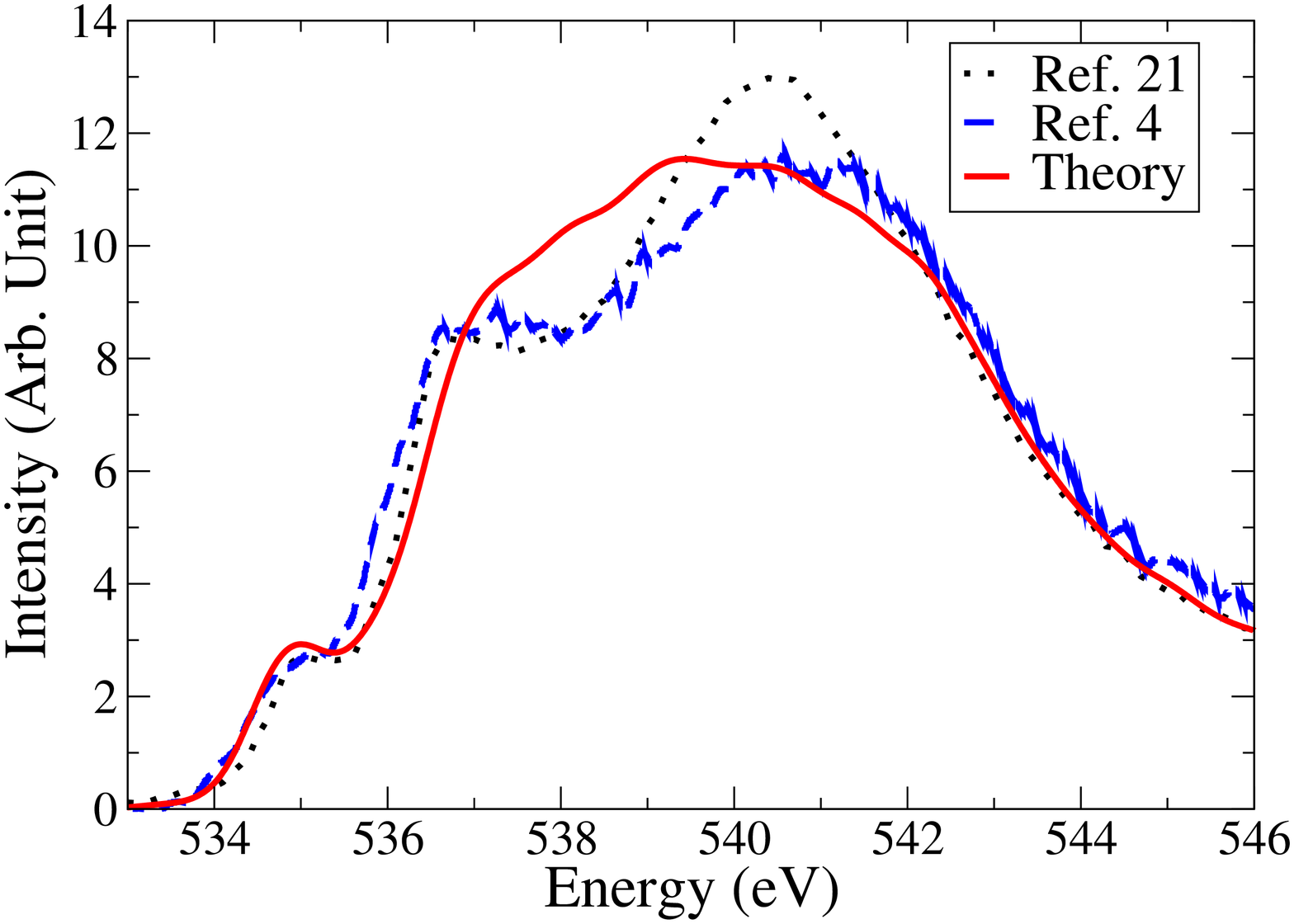,height=6cm,angle=270,clip=true}
 \epsfig{file=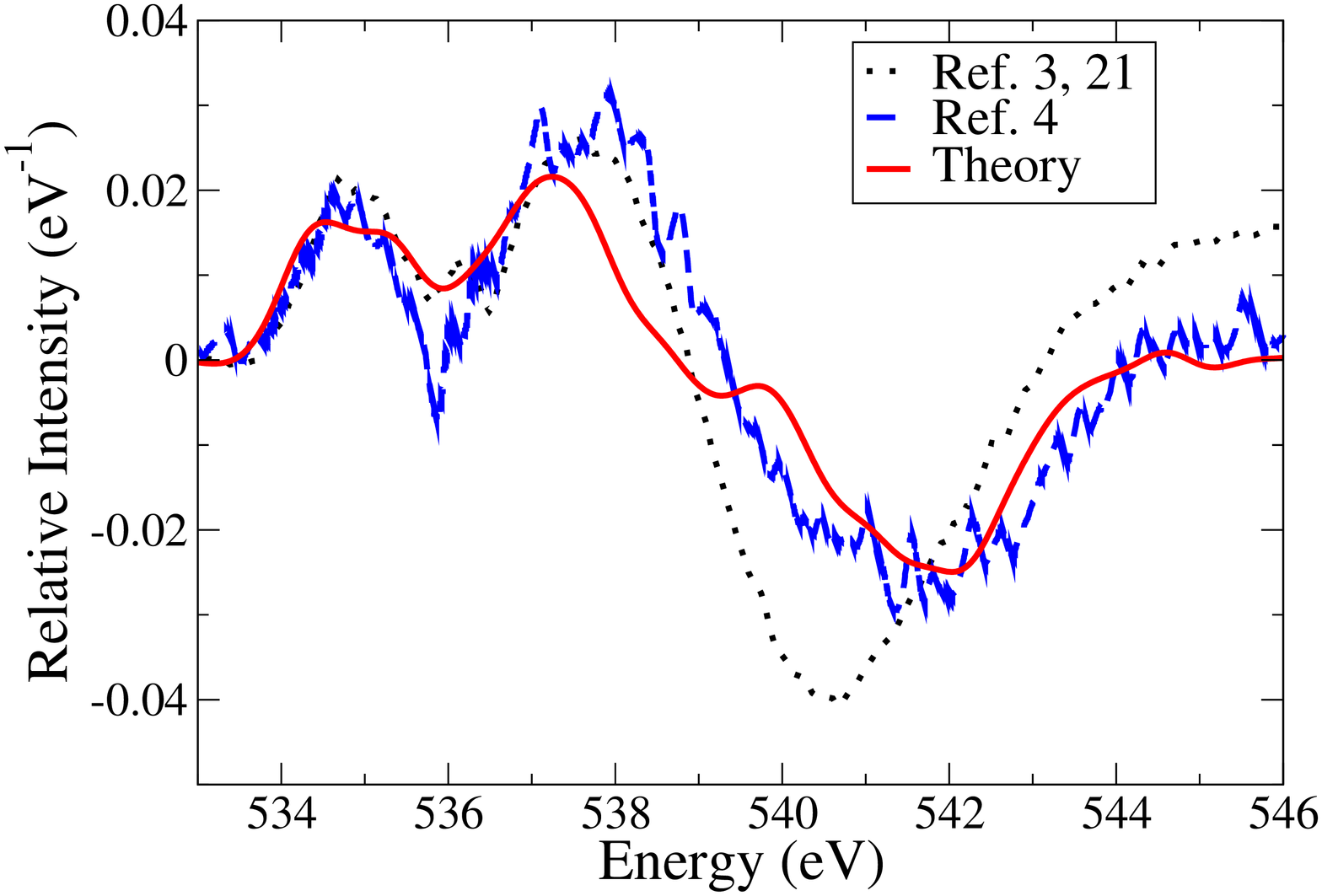,height=6cm,angle=270,clip=true}}
\caption{ (Color online) Comparison of calculated XAS spectra with experiments for water (top), ice Ih (middle) and
the difference between water and ice Ih (bottom). Two sets of 
experimental data are included for each system~\cite{Wernet2004, Nilsson2005, Tse2008}. Water is at room
temperature in theory and experiments, ice is at $T=269K$ in the present calculation, at $T=130K$ 
in~\cite{Nilsson2005} and at $T=40K$ in~\cite{Tse2008}. The difference spectrum is obtained from
[I(water)-I(Ih)]/S where I denotes spectral intensity and S is the integrated area of the spectra in the
energy range of the figure.}
\label{fig:Th_expr}
\end{figure}

In Fig.~\ref{fig:Th_expr} our results with quantum nuclei and COHSEX with local fields, are compared to XAS 
and XRS experiments.  The spectra are aligned at the onset and normalized by the same area.
The difference spectrum in the bottom panel of the figure is obtained by subtracting 
ice from water intensity dividing the result by the integrated spectral area.
The accord among experiments is excellent: the only noticeable difference between XAS and 
XRS in the post-edge of ice may reflect differences 
in the experiments and the samples. Overall the agreement 
between theory and experiment is quantitatively good in both water and ice. In the latter case, the dramatic 
improvement relative to Ref.~\cite{Wei2010} is a strong indication of the importance of  
a realistic model of disorder not only in liquid but also in crystalline water. The agreement between theory 
and experiments in the pre-edge is particularly satisfying given that most previous calculations failed to 
reproduce the correct relative intensity of this feature, associated to a bound exciton mostly localized 
on the excited molecule~\cite{Wei2010,Nordlund2007}. Based on the quantum-classical trends in Fig.~\ref{fig:xas_theory} 
we should expect that isotope effects should be detectable: making the protons more classical should 
weaken the pre-edge and strengthen the post-edge. Remarkably, this has been observed in high resolution XAS 
measurements on liquid H$_2$O and D$_2$O samples~\cite{Bergmann2007}. The main remaining difference between 
theory and experiment in the water spectrum is a small overestimate of the post-edge 
intensity. This may reflect residual overstructuring present in the H-bond network of the PI-AIMD 
simulations, a conclusion also borne by the analysis of the radial distribution functions~\cite{Morrone2008}.
We think that this inaccuracy is due to limitations of the BLYP functional approximation. Improved 
approximations including hybrid exchange that reduces the self-interaction error~\cite{Yang2008} and including 
dispersion (van der Waals) interactions would further soften the network structure~\cite{Zhaofeng2012} weakening 
the post-edge feature. The difference between ice and water in the relative strength of near- and post-edge 
features has its origin in the partial collapse of the H-bond network that brings a non-bonded molecular 
fraction within range of the first coordination shell in the liquid~\cite{Wei2010}. The corresponding ``interstitial" 
molecules do not form H-bonds with the central molecule but are H-bonded to molecules outside the coordination 
shell. This interpretation is consistent with diffraction studies~\cite{Finney2002} and we confirm it here. 
Quantum nuclei are crucial in water because they enhance the network collapse
compared to classical simulations. This effect is absent in ice Ih, which has an intact tetrahedral network with 
coordination 4~\cite{Interstitial_molecule}. In ice the residual differences between theory and experiments in 
the main edge may reflect to some extent the different temperatures of the simulation and the experiments. However, 
thermal effects should be small as the vibrational spectrum is largely ground-state dominated~\cite{Lin2011}. 
We think that the differences should mostly reflect the adopted screening approximation, 
which underestimates local fields. A more accurate screening model should enhance post-edge at the 
expense of near-edge, improving the agreement with experiment. A fully first-principles treatment of 
static screening would be possible but computationally very demanding. Finally, we have neglected the energy dependence 
of the self energy, which should affect the calculated spectra, particularly at high energy. Such effects could 
be included either via a dynamic COHSEX calculation or, simply, with phenomenological schemes such as 
rescaling the energy axis or using an energy-dependent broadening function~\cite{Materlik1983,Rehr2006}. 
The good agreement between experiment and theory found in the present study suggests that these effects should 
be small in water and ice.

In conclusion, we have improved an approach to calculate the XAS spectra of disordered environments~\cite{Wei2010},
 by including local field effects beyond the uniform screening approximation. The most important result of the 
present investigation has been, however, to elucidate the role of nuclear quantum dynamics on the spectra 
of water and ice. On one hand zero point motion broadens the spectra via Franck-Condon factors that we calculate
 by averaging over the distribution of the Feynman paths. On the other hand quantum nuclei modify the liquid 
structure by facilitating H-bond breaking. Our analysis shows that many competing factors are important to 
reproduce the spectral differences observed between water and ice. In view of its ability to 
probe local environments, x-ray spectroscopy is a very important tool for investigating not only bulk neat water,
 as done here, but also water solutions, water at interfaces, and, in general, soft condensed matter environments
 like bio-molecular solutions. The theoretical techniques developed here should contribute to these studies by 
giving insight on the origin of the spectral features and by allowing us to connect these features to the details of the local
molecular structure.

The authors thank J. A. Morrone for providing the path integral structures and 
acknowledge fruitful and stimulating discussions with L.G.M. Pettersson and A. Nilsson.
This work was supported by the 
NSF under grant CHE-0956500 and by the U.S. Department of Energy under grants DE-SC0005180 and DE-FG02-05ER4201. 
The calculations were performed at the National Energy Research Scientific Computing Center, 
which is supported by the Department of Energy under Contract DE-AC02-05CH11231.


\begin{thebibliography}{29}
\bibitem[{Int()}]{Interstitial_molecule}
\bibinfo{note}{Structures in which tetrahedrally bonded molecules penetrate the
  first coordination shell without forming a bond with the central molecule are
  found in high-pressure ice phases, an extreme case being that of ice VIII and
  VII.}
\end{thebibliography}

\end{document}